\documentclass[reprint,
superscriptaddress,
 amsmath,amssymb,
 aps,
prl,
raggedbottom,
]{revtex4-2}
\usepackage{graphicx}
\usepackage{siunitx}
\DeclareSIUnit\angstrom{\text {Å}}
\DeclareSIUnit\rydberg{Ry}
\usepackage{booktabs}
\usepackage{microtype}
\usepackage[labelfont=bf,justification=Justified]{caption}
\DeclareCaptionLabelSeparator{pipe}{~\textbar~}
\renewcommand{\thetable}{\arabic{table}}
\DeclareRobustCommand{\figref}[1]{{Figure~\ref{#1}}}
\DeclareRobustCommand{\tabref}[1]{{Table~\ref{#1}}}


\begin{document}

\preprint{APS/123-QED}

\title{Why ice is so slippery
}
\author{Sigbjørn Løland Bore}
\email{s.l.bore@kjemi.uio.no}
\affiliation{Hylleraas Centre for Quantum Molecular Sciences, Department of Chemistry, University of Oslo, PO Box 1033 Blindern, 0315 Oslo, Norway}

\author{B.N.J. Persson}
\email{b.persson@fz-juelich.de}
\affiliation{State Key Laboratory of Solid Lubrication, Lanzhou Institute of Chemical Physics, Chinese Academy of Sciences, 730000 Lanzhou, China}
\affiliation{Peter Gr\"unberg Institute (PGI-1), Forschungszentrum J\"ulich, 52425, J\"ulich, Germany}
\affiliation{MultiscaleConsulting, Wolfshovener str. 2, 52428 J\"ulich, Germany}

\author{Henrik Andersen Sveinsson}\email{h.a.sveinsson@fys.uio.no}
\affiliation{The Njord Centre, Department of Physics, University of Oslo, PO Box 1048 Blindern, 0316 Oslo, Norway}

\date{\today}
\begin{abstract}
We simulate nanoscale ice--glass friction from first principles and upscale it by modelling frictional heating. Nanoscale simulations alone overestimate friction and miss its velocity dependence. Frictional heating warms the contact toward the melting point, thickening the premelting film tenfold and cutting its viscosity a hundredfold; friction drops with velocity, matching experiment. Premelting forms the film without sliding or heating, but its slipperiness hinges on frictional heating, as Bowden and Hughes proposed in 1939, without melting. Snow, other materials and coatings remain open.
\end{abstract}

 \maketitle
The interest in ice friction has ancient roots, with early human activity involving sliding on ice from around 10000 B.C.\ using sleds \cite{gramly2024guide}. On this historical timescale, it was only very recently that the first scientific explanation for ice slipperiness was put forward in the 1800s by Michael Faraday's observations of premelting~\cite{faraday1859xxiv,jellinek1967liquid}: a liquid premelting film spontaneously forms on ice and provides lubrication. This explanation was followed by John Joly's pressure-melting theory~\cite{joly1887phenomena}: pressure causes a melting point depression, creating a lubricating premelting film. Subsequently, Bowden and Hughes proposed that the low friction of ice results from frictional heating \cite{bowden1939mechanism}: friction between a moving object on ice, such as a skate blade, generates heat, creating a thin film of liquid-like water on the ice surface. This water film acts as a lubricant, significantly reducing friction and enabling the smooth gliding motion characteristic of ice skating.

Modern science continues to debate the origin of ice friction. On the experimental side, Canale {\it et al.}\ emphasized the nanorheology of ice friction, measuring a visco-elastic interfacial film with a viscosity up to two orders of magnitude larger than that of bulk water and a rheology characteristic of a dense suspension~\cite{canale2019nanorheology}. Weber {\it et al.}\ combined friction experiments with MD simulations, attributing ice slipperiness to highly mobile water molecules at the ice surface~\cite{weber2018molecular}. Liefferink {\it et al.}\ followed up, investigating the roles of temperature, pressure, and velocity, proposing a ploughing mechanism~\cite{liefferink2021friction}. They concluded that the friction coefficient is not very sensitive to frictional heating. On the theoretical side, advances in computational efficiency have enabled molecular dynamics (MD) studies of nanoscale ice friction. Using force-field MD simulations, Baran {\it et al.}\ found the premelting film to be a robust phenomenon, reporting shear thinning and large slip lengths at hydrophobic interfaces as crucial for low ice friction, arguing for a minor role of frictional heating due to low film viscosity~\cite{baran2022ice,baran2024confinement}. More recently, At\i la {\it et al.}\ performed ice sliding on ice MD simulations, proposing a mechanism for displacement-driven amorphization at low temperatures~\cite{atila2025cold}. Their MD shows that a relatively thick (\SI{10}{nm}) liquid-like film can form at ice interfaces during slip with negligible temperature rise, through a competition between shear-induced amorphization and recrystallization. When sliding stops, the film recrystallizes, leaving only a few interfacial monolayers in non-ice configurations set by the counter-surface. Ice premelting at the ice--vapour interface has been observed in many experiments, but whether it also occurs at interfaces between ice and solids, in particular hydrophilic ones, is less well established. From these conflicting views, it can be concluded that no consensus has been reached on \emph{why ice is so slippery}.

\begin{figure*}[!ht]
    \centering
\includegraphics[width=1\linewidth]{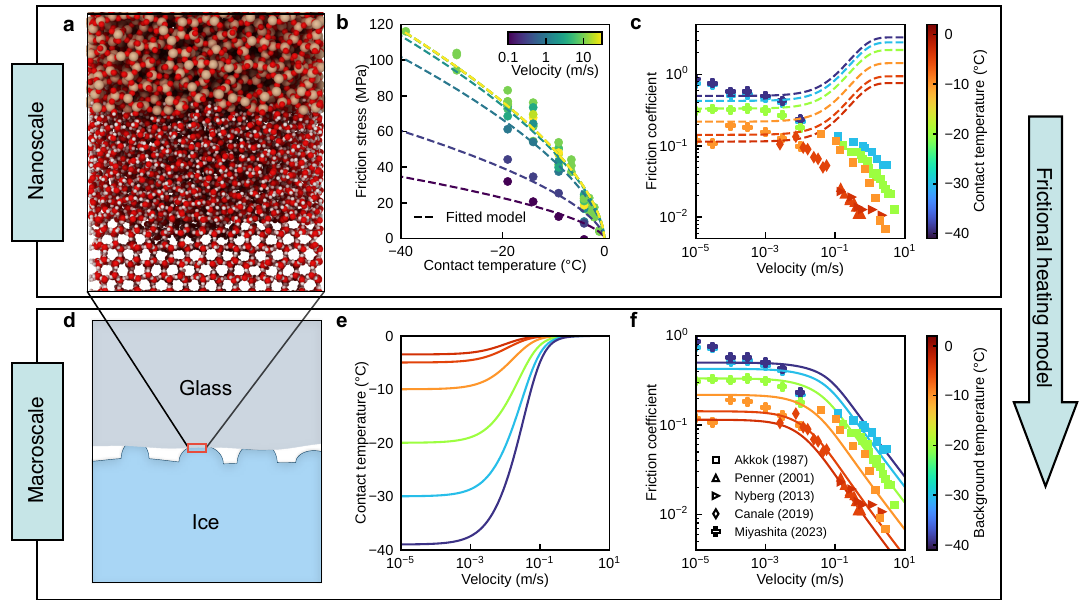}
    \caption{\textbf{Nanoscale vs.\ macroscale ice friction.} \textbf{a} Setup for nanoscale MD simulations of amorphous silica sliding on ice. \textbf{b} Friction stress as a function of contact temperature for different sliding velocities from nanoscale MD simulations, with fitted equation~\eqref{eq:high_friction} (dashed lines). \textbf{c} Ice friction coefficient as a function of velocity for different temperatures with equation~\eqref{eq:high_friction} directly, without frictional heating, comparing with experiments. \textbf{d} Illustration of macroscale friction, where nanoscale represents one asperity contact. \textbf{e} Contact temperature as a function of velocity for different background temperatures using the frictional heating model (equation~\eqref{eq:surface_temperature}). \textbf{f} Macroscale ice friction coefficient using the frictional heating model, comparing with experiments ($R^2_{\rm log} = 0.94$ on the full pool, $n=88$; $R^2_{\rm log} = 0.88$ on the held-out validation subset at $v \ge \SI{5e-3}{\metre\per\second}$, $n=55$, four sources --- see Methods). Contact temperatures for nanoscale MD simulations are reported in \si{\celsius} relative to the melting point of the model concerned: \SI{289}{\kelvin} for DC-R$^2$SCAN and \SI{270}{\kelvin} for ClayFF/TIP4P-Ice. Experimental friction coefficient data are from ice--glass friction experiments~\cite{canale2019nanorheology,miyashita2023sliding,akkokParametersAffectingKinetic1987} and curling stone experiments~\cite{penner2001physics,nyberg2013asymmetrical}, whose stones are composed of granite with a silica content of $\sim$\SI{70}{\percent}.}
    \label{fig:upscaling}
\end{figure*}

In this paper, we present evidence that, at rock-like counterfaces, a premelting film forms without sliding or heating, and frictional heating is the primary cause of ice slipperiness. Applying machine-learning interatomic potentials (MLIPs) based MD, we simulate nanoscale ice friction against a realistic ice--glass (amorphous silica) interface---a model system for ice--rock contacts in nature---from first principles (see \figref{fig:upscaling}a and the Supplemental Material for details). By integrating our nanoscale MD findings into the frictional heating model developed by Persson~\cite{persson2015ice}, we predict the macroscopic ice friction behavior. Comparing our macroscale predictions with experimental ice--glass friction data, we find excellent agreement, supporting the frictional heating theory of Bowden and Hughes.

The key control parameters in ice friction experiments are the background temperature $T_0$ of the ice and the velocity $v$ of the sliding body. Direct prediction of experimental ice friction conditions using MD is impossible due to the large length and time scales involved. However, from nanoscale MD simulations, we can obtain the frictional shear stress $\sigma_{\rm f}(v,T)$ acting in the asperity contact regions for given sliding velocity $v$ and contact temperature $T$ (see the Supplemental Material).
We perform such simulations of the ice--glass contact (\figref{fig:upscaling}a), for $v\gtrsim \SI{0.1}{\metre\per\second}$, with the friction stress shown in \figref{fig:upscaling}b. The friction stress from nanoscale MD is well-modeled by the following expression:
\begin{equation}
    \sigma_{\rm f}(v,T) = \sigma_{\rm k}\left(1-\frac{T}{T_{\rm m}}\right)^{\beta}\left[1-\left(1-\frac{\sigma_{\rm c}}{\sigma_{\rm k}}\right)e^{-\tfrac{v}{v_{\rm k}}}\right],
    \label{eq:high_friction}
    \end{equation}
Equation~\eqref{eq:high_friction} can be rationalized as a sigmoid between two limiting behaviors:
\begin{align}
    \sigma_{\rm f}(v,T) = \sigma_{\rm c} \left(1-\dfrac {T}{T_{\rm m}}\right)^{\beta}, \quad v \ll v_{\rm k}, \label{eq:friction-stress-creep}\\
    \sigma_{\rm f}(v,T) = \sigma_{\rm k} \left(1-\dfrac {T}{T_{\rm m}}\right)^{\beta}, \quad v \gg v_{\rm k},
\end{align}
representing lower and upper velocity limits, respectively. Here, $\sigma_{\rm k} = \SI{391}{\mega\pascal}$, $v_{\rm k} = \SI{0.53}{\metre\per\second}$ and $\beta = 0.61$ are fit to our nanoscale MD friction-stress data for \SIrange{0.1}{31.6}{\metre\per\second}, and $\sigma_{\rm c} = \SI{59}{\mega\pascal}$ is anchored to the low-velocity ice--glass friction-coefficient data of Miyashita~\textit{et al.}~\cite{miyashita2023sliding} at $v < \SI{5e-3}{\metre\per\second}$, where equation~\eqref{eq:high_friction} reduces to equation~\eqref{eq:friction-stress-creep} within $\sim 5\%$. The four parameters are determined by an iterative co-fit (Methods). Equation~\eqref{eq:friction-stress-creep} has previously been proposed by Persson for ice--ice friction experiments for low velocities~\cite{persson2015ice}, albeit with $\beta=0.15$.

To incorporate heat transport effects, we use a frictional heating model. By assuming that ice and glass make contact at asperities, making up only a small fraction of the apparent contact area (\figref{fig:upscaling}d), the contact temperature can be expressed as~\cite{green, persson2015ice} (see Methods below for additional details):
\begin{equation}
    T = T_0 + \sigma_{\rm f}(v,T)\frac{s(v)v R_{\rm p}}{\kappa} \left(1 + \frac{\pi}{8}\frac{R_{\rm p} v}{D}\right)^{-\frac 12}.
    \label{eq:surface_temperature}
\end{equation}
Here, $R_{\rm p}$ is the asperity contact radius, $D$ is the thermal diffusivity, $\kappa$ is the thermal conductivity, all of which are available from literature data (see Table~\ref{tab:parameters}). $s(v)$ is the proportion of heat that is absorbed by the slider, and can be explicitly expressed as a function of the sliding velocity (see Methods below), typically in the range from 0.6 to 1 with increasing sliding velocity. By inserting equation~\eqref{eq:high_friction} into equation~\eqref{eq:surface_temperature}, we obtain an implicit equation for the contact temperature at a given background temperature and sliding velocity, which we solve iteratively to obtain $T$ and $\sigma_{\rm f}$. Of the four parameters in equation~\eqref{eq:high_friction}, only $\sigma_{\rm c}$ is anchored to experiment---specifically, Miyashita~\textit{et al.}~\cite{miyashita2023sliding} at $v < \SI{5e-3}{\metre\per\second}$; $\sigma_{\rm k}$, $v_{\rm k}$, and $\beta$ are fit to our nanoscale MD simulations (see Methods below). The friction coefficient can then be estimated from friction stress as $\mu = \sigma_{\rm f}/\sigma_{\rm P}$, by assuming contact pressure equals ice penetration hardness, with $\sigma_{\rm P}\simeq \SI{35}{\mega\pascal}$ (see Methods below).

\figref{fig:upscaling}c reports the nanoscale ice friction coefficient as a function of sliding velocity for different temperatures, using the nanoscale MD results directly. When compared with experimental ice friction data, the coefficient of friction is not only grossly overestimated in magnitude but also shows the wrong trend: friction increasing with sliding velocity, in opposition to experiments. Using the frictional heating model described by equation~\eqref{eq:surface_temperature}, we can upscale nanoscale MD results to the macroscale. The resulting contact temperature is reported in \figref{fig:upscaling}e. At low velocities, frictional heating is negligible, with the contact temperature close to the background temperature. At higher velocities, frictional heating increases the contact temperature, which approaches the melting point for a sliding velocity of around \SI{0.1}{\metre\per\second}. This change in contact temperature dramatically changes the ice friction coefficient behavior from the nanoscale ice friction coefficients in \figref{fig:upscaling}c to the macroscale ice friction coefficients in \figref{fig:upscaling}f, which shows excellent agreement with experimental data \cite{akkokParametersAffectingKinetic1987, penner2001physics, nyberg2013asymmetrical, canale2019nanorheology, miyashita2023sliding}, having an $R^2_{\log}$ coefficient of determination of \num{0.94} over the full experimental pool ($n=88$), and \num{0.88} on a held-out validation subset of $n=55$ points from four sources (Akkok, Penner, Nyberg, Canale) at $v \ge \SI{5e-3}{\metre\per\second}$, where the macroscale prediction is independent of the $\sigma_{\rm c}$ anchor (see Methods).

\begin{figure}[!htb]
    \centering
\includegraphics[width=1\linewidth]{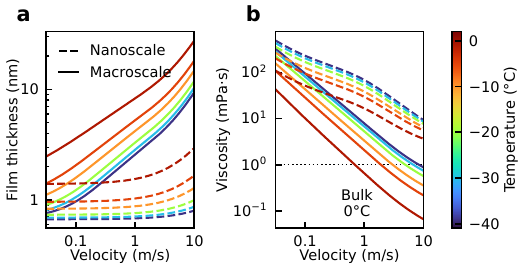}
    \caption{\textbf{Nanoscale vs.\ macroscale premelting film properties.} \textbf{a} premelting film thickness and \textbf{b} viscosity estimated directly from nanoscale MD simulations, and to macroscale using the frictional heating model. Nanoscale is reported for contact temperature, while macroscale is reported for background temperature.
    }
    \label{fig:film-properties}
\end{figure}

From the nanoscale MD simulations, we can extract properties of the premelting film, which we can also upscale to the macroscale using the frictional heating model. In \figref{fig:film-properties}a, we report the premelting film thickness as a function of sliding velocity. The premelting film thickness increases with sliding velocity at the nanoscale, but the effect is significantly amplified at the macroscale, rising by about an order of magnitude. At fixed contact temperature the film thickens by only a factor of \numrange{2.4}{3.8} across the simulated velocity range, \SIrange{0.1}{31.6}{\metre\per\second}: shear-induced disordering advances with sliding distance~\cite{pastewka2011diamond, moras2018shearmelting, atila2025cold}, whereas recrystallization proceeds thermally, at a rate set by temperature rather than by the sliding, so faster sliding disorders the interface faster, the recovery rate is unchanged, and the steady-state thickness grows with velocity. The much larger macroscale amplification is thermal: frictional heating drives the contact toward $T_{\rm m}$, where recrystallization slows steeply, and at \SI{10}{\metre\per\second}, when the contact temperature goes from \SIrange{-19}{-1}{\celsius}, the film thickens by a further factor of $\sim$3. This amplification is also evident when considering viscosity, which drops by about two orders of magnitude. It is these two effects combined that cause the strong drop in ice friction with increasing sliding velocity at the macroscale.

\begin{figure}[!htb]
    \includegraphics[width=1\linewidth]{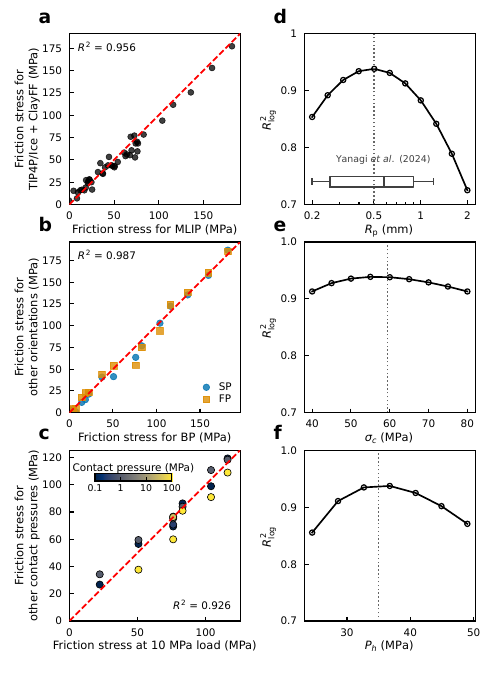}
\caption{\textbf{Robustness of the nano-to-macro friction pipeline.}
\textbf{a} Parity of friction stress between the DC-R$^2$SCAN MLIP and the
classical TIP4P/Ice + ClayFF force field at matched basal-plane (BP)
configurations, paired by velocity and supercooling $\Delta T$ from the
respective model melting point.
\textbf{b} Parity of friction stress at the basal plane against the
secondary prismatic (SP) and first prismatic (FP) at \SI{10}{\mega\pascal} contact pressure,
matched by temperature and sliding velocity.
\textbf{c} Parity of friction stress at \SI{10}{\mega\pascal} contact pressure against
friction stress at other contact pressures spanning \SIrange{0.1}{100}{\mega\pascal},
basal plane only.
\textbf{d}--\textbf{f} Log-space coefficient of determination
$R^2_\mathrm{log}$ of the macroscale friction prediction against the
pooled experimental dataset, as the contact radius $R_{\rm p}$ (\textbf{d}), the
friction-stress floor $\sigma_{\rm c}$ (\textbf{e}), and the penetration hardness $\sigma_{\rm P}$ (\textbf{f})
are varied; vertical dotted lines mark the values adopted in the main
analysis. The horizontal box-whisker in \textbf{d} shows experimental
ice pebble radius range reported by
Yanagi~\textit{et al.}~\cite{yanagi2024characteristics}.}
    \label{fig:robustness}
\end{figure}

The validity of our findings rests on several factors, whose robustness we quantify in \figref{fig:robustness}. First, the friction stress at the nanoscale rests on the atomistic model, which is DC-R$^2$SCAN MLIP in our case. In the Supplemental Material, we quantify the validity of the DC-R$^2$SCAN MLIP model, showing excellent agreement with DFT reference and representation of physical properties of water, ice and amorphous silica. However, a single model may be biased, and the modeling of water using DFT is notoriously difficult~\cite{gillan2016perspective}. Therefore, we repeated the sweep with force-field simulations using TIP4P/Ice for water and ClayFF for the silica (see the Supplemental Material for details). As demonstrated by the parity plot in \figref{fig:robustness}a, the force-field-based simulations are in very close agreement with the MLIP, with an $R^2$ of \num{0.96}, indicating that nanoscale friction behavior is not highly model-sensitive. We also computed the friction stress for different orientations of ice planes, first prismatic and second prismatic (FP and SP), which are known to have different premelting behaviors~\cite{conde2008thickness,espinosa2016ice}. For these, we get an $R^2$ of \num{0.99}, which is within the statistical noise of sampling. Finally, for contact pressure, there are differences, especially at high temperatures and high contact pressures beyond the penetration hardness (for which structures can fully melt), but for the most part friction stress of the MLIP is largely insensitive to contact pressure, with an $R^2$ of \num{0.93}, which is in line with contact pressure not playing a major role, except when beyond the penetration hardness. We note that film thickness is more sensitive to model, orientation and contact pressure than friction stress, as shown in \figref{sm-fig:thickness-parity} of the Supplemental Material: the force-field films are thicker than the MLIP ones by a median factor of \num{1.6}, and the ratio increases with sliding velocity, even though the friction stresses agree closely. The $R^2_\mathrm{log}$ of our estimated coefficient of friction against experiments upscaled from nanoscale friction stress, rests on several assumptions, of which we report the most sensitive ones in \figref{fig:robustness}d, e and f. It is most sensitive to the radius of the ice pebbles, but within the range of experimentally determined values \cite{yanagi2024characteristics}, we have 0.8 agreement with experiment. $\sigma_{\rm c}$ and $\sigma_{\rm P}$ affect $R^2_{\log}$ in a similar manner, scaling the predicted coefficient of friction for low velocities (see equation~\eqref{eq:friction-stress-creep}) and all velocities, respectively. Other parameters have smaller effects, as shown in \figref{sm-fig:extended-sensitivity}.

Our findings provide a window into reconciling conflicting views on ice friction present in the literature. Baran \textit{et al.}~\cite{baran2022ice,baran2024confinement} highlight shear thinning as important for ice friction. We also observe it in our nanoscale simulations (\figref{fig:film-properties}b). In zero-velocity simulations, premelting alone establishes the initial film between the ice and glass surfaces, and this film holds near \SI{5.5}{\angstrom} down to \SI{-89}{\celsius} and thickens to \SIrange{10}{11}{\angstrom} within \SI{1}{\celsius} of melting in both models (\figref{sm-fig:thickness-parity}e--f). At{\i}la {\it et al.}~\cite{atila2025cold} have shown that at ice--ice interfaces, the film is instead established by displacement-driven amorphization. Which route dominates is set by the counterface: at the hydrophilic ice--glass interface studied here, premelting alone maintains a film at rest, whereas at their ice--ice interface the film recrystallizes when sliding stops, leaving only a few interfacial monolayers, so there the sliding itself must generate the film. Displacement-driven amorphization likely plays a role in thickening our film, which grows by a factor of \numrange{2.4}{3.8} with sliding velocity at fixed contact temperature (\figref{fig:film-properties}a). That the two routes lead to a similar state is supported structurally: At{\i}la {\it et al.}\ characterise their shear-amorphized ice as very close to supercooled water, reporting only minor deviations in the pair distribution function (their SM), and our steady-state O--O $g(r)$ inside the premelting film matches this picture in both models and at every $(T, v)$ probed (\figref{sm-fig:strip-rdf}). The two pictures are therefore complementary rather than competing; frictional heating governs the friction once the film is established. At{\i}la {\it et al.}\ anticipate this, noting that recrystallization slows rapidly as the melting point is approached, so that the film thickens, and that the frictional heating argument may still hold, ``though not to the extent that liquefaction should be attributed to thermal melting''. Our model describes frictional heating without thermal melting: it contains no melting enthalpy, and the friction stress in equation~\eqref{eq:high_friction} decreases continuously as $(1-T/T_{\rm m})^{\beta}$ without a phase transition. The film properties also depend on crystal orientation~\cite{conde2008thickness,espinosa2016ice} (\figref{fig:robustness}b). However, even with these effects present, we have shown that nanoscale simulations do not capture the velocity dependence of ice friction (\figref{fig:upscaling}c), and that frictional heating is a necessary ingredient to describe the observed velocity dependence qualitatively and quantitatively. Finally, pressure melting is already considered unlikely in most situations, as ice asperity contact regions yield by plastic flow before the pressure becomes high enough to form liquid water, and our results corroborate that this is an unnecessary hypothesis to explain the slipperiness of ice: the friction stress only weakly depends on contact pressure under the pressures probed (\SIrange{0.1}{100}{\mega\pascal}, \figref{fig:robustness}c).

Several aspects of ice friction remain to be explored. The predictions reported here apply to a specific ice--glass counterface, and not to ice-on-ice or ice-on-other-material configurations more generally; materials with higher thermal conductivity, such as metals, would lead to different thermal feedback. We have not considered in depth the effects of surface chemistry, such as hydroxylated or hydrophobic surfaces (e.g., waxes used in skiing), snow, ploughing~\cite{liefferink2021friction}, or the formation of ice-particle slurries. Canale {\it et al.}~\cite{canale2019nanorheology} report a thicker (order $\sim$\SI{100}{\nano\metre}), more viscous interfacial film whose viscosity rise toward the melting point reflects this thick-film regime rather than the velocity dependence of friction we address. We note, however, that the thickness of even ordinary premelting layers remains contested across experimental techniques, with simulations agreeing more closely~\cite{slater2019premelting}. Nevertheless, the velocity dependence of the friction coefficient data by Canale {\it et al.}\ (measured for ice against a borosilicate-glass bead, an ice--glass contact) is consistent with our predictions (\figref{fig:upscaling}f). Our macroscale predictions are anchored to experiment at one point: the low-velocity friction floor $\sigma_{\rm c}$ is anchored to Miyashita~\textit{et al.}~\cite{miyashita2023sliding} at $v < \SI{5e-3}{\metre\per\second}$, so the agreement at very low velocities is by construction rather than a prediction; agreement at $v \ge \SI{5e-3}{\metre\per\second}$, which spans four other experimental sources, is a genuine held-out prediction. The Persson heat model treats the slider as making contact at asperities of a single radius (here $R_{\rm p} = \SI{0.5}{\milli\metre}$, within the experimental range reported in Ref.~\cite{yanagi2024characteristics}); a real ice slider carries a distribution of asperity sizes, with the associated surface roughness and rms slope, that we do not convolve over. The melting point of our DC-R$^2$SCAN MLIP is $T_{\rm m} = \SI{289}{\kelvin}$, about \SI{16}{\kelvin} above the experimental value, so absolute contact temperatures here should be read relative to $T_{\rm m}$. These are promising directions for future work, requiring additional simulations and different theoretical frameworks.

Our analysis focuses on the velocity dependence of ice friction, but as emphasized by Rosenberg, ice is slippery even while standing still~\cite{rosenberg2005ice}, potentially at odds with frictional heating hypotheses. For the contacts studied here, this paradox is resolved by the picture we provide. A premelting film is already present at rest, so the interface is lubricated before anything moves. Premelting, not frictional heating, forms this film. Frictional heating then makes the ice far more slippery at very modest velocities: movements sustained over the scale of one millimeter, with tangential velocities of $\sim$\SI{0.1}{\metre\per\second}, warm the contact and lead to dramatic drops in friction. This is qualitatively consistent with everyday experience: slow, careful steps below \SI{0.1}{\metre\per\second} retain grip, but rapid scuffing to regain grip ends in catastrophe.

\input{manuscript.bbl}
\section{Methods}
\subsection{Frictional heating modeling}
We determine the contact temperature at the ice surface and the slider due to frictional heating as follows. We assume that the slider is flat and that it makes contact with the ice substrate only at isolated pebble-shaped asperities of radius $R_{\rm p}$. These asperities make up only a small proportion of the apparent contact area, as illustrated in \figref{fig:upscaling}d. Due to frictional heating, there will be an increase $\Delta T$ of the contact temperature $T$ with respect to the background temperature $T_0$. Seen from the slider, the contacts with the ice pebbles of the substrate are disk-shaped moving heat sources that provide a heat flux $J$. The maximum temperature increase within the contact region is, in this case, very well approximated by equation (8) in Ref.~\cite{green}:
\begin{equation}
    \Delta T = \frac{s(v) J R_{\rm p}}{\kappa_{\rm slider}} \left(1+\frac{\pi}{8}\,\frac{R_{\rm p} v}{D_{\rm slider}}\right)^{-1/2},
    \label{eq:deltaT}
\end{equation}
$D$ is the thermal diffusivity, $\kappa$ is the thermal conductivity, and $s$ is the proportion of the heat that goes into the slider (the rest, $1-s$, goes into the ice). This expression includes both the effect of normal heat transfer away from the contact and the effect of new cold glass constantly entering the contact region. It is useful to write equation~\eqref{eq:deltaT} as
\begin{equation*}
    s(v)J = \alpha_{\rm slider}\Delta T
\end{equation*}
with
\begin{equation*}
    \alpha_{\rm slider} = \frac{\kappa_{\rm slider}}{R_{\rm p}} \left(1+\frac{\pi}{8}\,\frac{R_{\rm p} v}{D_{\rm slider}}\right)^{1/2}.
\end{equation*}

For the substrate, with stationary ice pebbles, the maximum temperature elevation is \cite{green}:
\begin{equation*}
    \Delta T = \frac{(1-s(v))J R_{\rm p}}{\kappa_{\rm substrate}},
\end{equation*}
or equivalently
\begin{equation*}
    (1-s(v)) J = \alpha_{\rm substrate}\Delta T,
\end{equation*}
with
\begin{equation*}
\alpha_{\rm substrate} = \frac{\kappa_{\rm substrate}}{R_{\rm p}}.
\end{equation*}
We approximate the temperatures on the ice surface and the silica glass surface to be equal (i.e., the contact temperature)~\cite{HeatTransfer}. We thus obtain the partitioning coefficient of the heat flux:
\begin{equation}
    s(v) = \frac{\alpha_{\rm slider}}{\alpha_{\rm substrate} + \alpha_{\rm slider}}.
\end{equation}
This partitioning tends towards 0.5 for stationary contacts between materials with similar thermal properties, whereas it tends towards 1 at high sliding velocities, meaning almost all the heat goes into the slider. Using the MD-derived friction law in equation~\eqref{eq:high_friction}, we insert $J=\sigma_{\rm f} v$ into equation~\eqref{eq:deltaT} to recover the contact temperature relation given in equation~\eqref{eq:surface_temperature}. Since $\sigma_{\rm f}$ is a function of the contact temperature, equation~\eqref{eq:surface_temperature} is an implicit equation for the contact temperature $T$ in the region. We solve it iteratively and then obtain $\sigma_{\rm f}$ from the same constitutive relations. Note that we do not consider squeeze-out of the premelting film, as for nanometer-thick films at macroscopic contact dimensions, squeeze-out is negligible even at the low viscosity of water. In addition, during sliding, shear-induced amorphization would continuously produce a liquid-like film on the ice surface~\cite{atila2025cold}.

To obtain friction stresses for comparison with experiments, we insert parameters taken from the curling literature, because typical curling ice has been very well characterized; it is pebbled with a contact diameter of around \SI{2}{\milli\metre} \cite{li2022experimental}. To obtain roughly the midpoint temperature instead of the maximum temperature of the contacts, we set $R=\SI{0.5}{\milli\metre}$. Curling stones have a very flat contact surface, which is consistent with the heat model described above. These stones are made of granite, which is mostly a silicate mineral, for which our amorphous silica glass is a reasonable analogue. In the frictional heating model, we use thermal properties of the ice and granite as listed in \tabref{tab:parameters}. We compute the thermal diffusivity needed in equation~\eqref{eq:surface_temperature} by $D=\kappa /(\rho C_{\rm P})$.
\begin{table}[!htb]
    \centering
        \caption{\textbf{Material properties used in the frictional heating model.} Granite properties were obtained from Ref.~\cite{mirandaEffectTemperatureThermal2019}, and ice properties from Ref.~\cite{hobbs1974ice}.}
    \begin{tabular}{lccc}
                \toprule
        Property & Unit & Ice & Granite \\
        \midrule
        Thermal conductivity $\kappa$ & (\si{\watt\per\metre\per\kelvin}) & 2.2 & 3.2 \\
        Density $\rho$ & (\si{\gram\per\centi\metre\cubed}) & 0.917 & 2.7 \\
        Specific heat $C_{\rm P}$ & (\si{\joule\per\kilo\gram\per\kelvin}) & 2100 & 800 \\
        \bottomrule
    \end{tabular}
    \label{tab:parameters}
\end{table}
\subsection{Friction coefficient}
Relating the frictional shear stress to the friction coefficient is nontrivial because it depends on whether the surfaces deform elastically (as expected for sufficiently smooth surfaces) or plastically. Plastic deformation of ice is a stress-aided, thermally activated process, and the penetration hardness of ice depends on the temperature and the indentation velocity. Here we note that if a slider is slid repeatedly on ice, the high ice asperities will be smoothed plastically during the first few sliding events, so after run-in, the surfaces may deform only elastically, and subject to a contact pressure on the order of the penetration hardness of ice of about $\sigma_{\rm P} = \SI{35}{\mega\pascal}$, collected from experimental data~\cite{hobbs1974ice,penner2001physics}. We then compute the friction coefficient $\mu = \sigma_{\rm f}/\sigma_{\rm P}$

\subsection{Ice--glass friction model parameterization}
\begin{figure}[!t]
    \centering
    \includegraphics[width=1\linewidth]{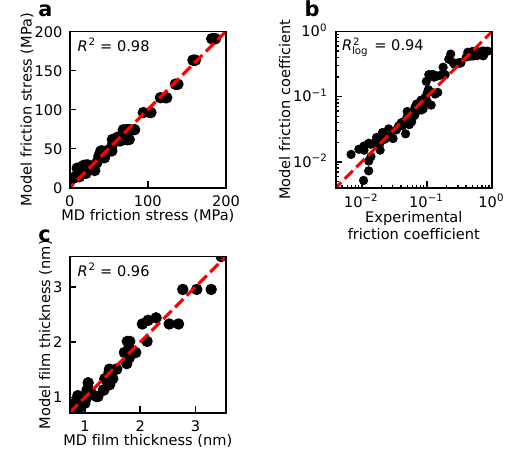}
    \caption{\textbf{Parity plots for ice--glass friction modeling vs.\ data}. \textbf{a} Nanoscale MD data vs.\ fitted model predictions for  friction stress equation~\eqref{eq:high_friction}, corresponding to \figref{fig:upscaling}b.  \textbf{b} Experimental vs.\ modeled macroscale friction coefficient, corresponding to \figref{fig:upscaling}f: all $n=88$ experimental points ($R^2_{\log}=0.94$; $R^2_{\log}=0.88$ on the $n=55$ held-out validation points).  \textbf{c} MD data vs.\ fitted model predictions for premelting film thickness, corresponding to \figref{fig:film-properties}a.}
    \label{fig:parity_plots}
\end{figure}

We represent our nanoscale MD data by analytical functions parameterized based on nanoscale ice--glass friction data at specific state points. The melting points of our water models differ from the experimental one: \SI{289}{\kelvin} for DC-R$^2$SCAN and \SI{270}{\kelvin} for ClayFF/TIP4P-Ice. We therefore report temperatures in \si{\celsius} relative to the melting point of the model concerned, to enable comparison with experimental data. For equation~\eqref{eq:high_friction}, the four parameters are determined by an iterative co-fit with a strict train--validate split. $\sigma_{\rm k}$, $v_{\rm k}$, and $\beta$ are anchored to nanoscale MD friction-stress data at \SI{10}{\mega\pascal} contact pressure; the low-velocity floor $\sigma_{\rm c}$ is anchored to ice--glass friction-coefficient data of Miyashita~\textit{et al.}~\cite{miyashita2023sliding} restricted to $v < \SI{5e-3}{\metre\per\second}$ ($n=28$ points), where equation~\eqref{eq:high_friction} reduces to equation~\eqref{eq:friction-stress-creep} within $\sim 5\%$. The iteration alternates (i)~refitting ($\sigma_{\rm k}$, $v_{\rm k}$, $\beta$) on the MD data with $\sigma_{\rm c}$ pinned and (ii)~refitting $\sigma_{\rm c}$ on the 28 Miyashita low-velocity points with ($\sigma_{\rm k}$, $v_{\rm k}$, $\beta$) pinned; it converges in two iterations to the values quoted above ($\sigma_{\rm c} = \SI{59}{\mega\pascal}$, $\sigma_{\rm k} = \SI{391}{\mega\pascal}$, $v_{\rm k} = \SI{0.53}{\metre\per\second}$, $\beta = 0.61$). The held-out validation pool used for the $R^2_{\log}=0.88$ reported in the main text consists of all macroscale experimental points from Akkok, Penner, Nyberg, and Canale at $v \ge \SI{5e-3}{\metre\per\second}$ ($n=55$); the training set and validation pool have zero source overlap and zero velocity-range overlap. Parity plots of how well the model predictions match the MD data are shown in \figref{fig:parity_plots}a, and how macroscale predictions match the experimental friction coefficient in \figref{fig:parity_plots}b. For the premelting film thickness, we fitted the following functional form for simulations with temperatures between 260 and \SI{289}{\kelvin}:
\begin{equation}
h(v,T)
= h_0
+ a\left[-\ln\!\left(1-\frac{T}{T_{\rm m}}\right)\right]^{\gamma}
\left(1+\frac{v}{v_0}\right).
\label{eq:film_thickness}
\end{equation}
The offset is held fixed at $h_0=\SI{0.60}{\nano\metre}$. The remaining parameters, $a=\SI{13.06}{\pico\metre}$, $\gamma=2.37$ and $v_0=\SI{5.06}{\metre\per\second}$, are fitted to MD data for premelting film thickness as a function of temperature and sliding velocity, with parity plot in~\figref{fig:parity_plots}c. We note that Limmer and Chandler proposed such a logarithmic dependence for temperature for premelting thickness in ice--air \cite{limmer2014premelting}, albeit without $\gamma$ exponent. A parity plot of how well the model predictions match the MD data is reported in \figref{fig:parity_plots}c. Equation~\eqref{eq:film_thickness} is plotted in \figref{fig:film-properties}a, with premelting film viscosity in \figref{fig:film-properties}b computed using $\eta = \sigma_{\rm f}h/v$.

\subsection{Data and code availability}
The code, derived data, and figures supporting this work are archived in an open data repository~\cite{data_repository}, \href{https://doi.org/10.5281/zenodo.20407640}{doi:10.5281/zenodo.20407640}. License: MIT (code), CC-BY-4.0 (data).

\section{Acknowledgments}
This work was supported by the Research Council of Norway through the Centre of Excellence Hylleraas Centre for Quantum Molecular Sciences (Grant 262695) and the Young Researcher Talent grants 344993 and 354100. We acknowledge the EuroHPC Joint Undertaking for awarding this project access to the EuroHPC supercomputer LUMI, hosted by CSC (Finland) and the LUMI consortium through a EuroHPC Regular Access call (Grants EHPC-REG-2023R02-088, EHPC-REG-2023R03-146). Support was also received from the Centre for Advanced Study in Oslo, Norway, which funded and hosted the SLB Young CAS Fellow research project during the academic years of 23/24 and 24/25. Part of the simulations were performed on resources provided by Sigma2 --- the Norwegian National Infrastructure for High-Performance Computing and Data Storage (grant numbers NN4654K and NS4654K). Generative AI (Claude Opus, Anthropic) was used to assist with manuscript editing and revision. All authors have reviewed and take full responsibility for the content.

\section*{Competing interests}
The authors declare no competing interests.

\setcounter{figure}{0}
\setcounter{table}{0}
\setcounter{equation}{0}
\renewcommand{\thefigure}{S\arabic{figure}}
\renewcommand{\thetable}{S\arabic{table}}
\renewcommand{\theequation}{S\arabic{equation}}

\section*{Supplemental Material}

\section{Models and simulation protocols}
This document details the molecular dynamics methodology used to generate the nanoscale ice friction data discussed in the main text.

\subsection{First-principles reference}
To model nanoscale ice--glass friction from first principles accurately, we describe the electronic structure using density-corrected density-functional theory (DC-DFT) with the R$^2$SCAN functional (DC-R$^2$SCAN). DC-DFT has been extensively benchmarked for water, showing not only excellent correspondence in energies with respect to coupled cluster calculations, but also for liquid water properties~\cite{dasgupta_elevating_2021,dasgupta2025nuclear,dasgupta2025reactivity}. The atomic core electrons were described using Goedecker--Teter--Hutter (GTH) pseudopotentials, while valence electron molecular orbitals were expanded in triple-zeta double-polarized basis sets (TZV2P) optimized for the SCAN functional. The kinetic energy cutoff for the plane-wave expansion of the density was set to \SI{2500}{\rydberg}, which was required to get a numerical accuracy of $\sim$\SI{2}{\milli\electronvolt\per\angstrom} for forces. The truncated Coulomb operator with a cutoff radius \SI{6}{\angstrom} was used, corresponding to approximately half the length of the smallest edge of the simulation cell. To overcome the computational cost of the Hartree--Fock calculation, the alternating direction method of multipliers (ADMM) approximation was used with an optimized valence basis set (BASIS\_ADMM\_UZH). The Schwarz integral screening threshold was set to $10^{-6}$ atomic units, starting from a converged PBE calculation.

\subsection{Machine-learning interatomic potential}
To enable nanoscale ice--glass friction simulations from first principles with DC-R$^2$SCAN at system sizes of thousands of atoms and timescales up to hundreds of nanoseconds, we trained computationally efficient machine-learning interatomic potentials (MLIPs) to reproduce DC-R$^2$SCAN energies and forces. As the MLIP, we used DeePMD~\cite{zeng_deepmd-kit_2023,zeng2025deepmd} with the se\_e2\_a architecture and 25, 50, or 100 neurons for the hidden embedding layer, and a 16-neuron submatrix of the embedding matrix. A distance cutoff of \SI{6}{\angstrom} was used with a smoothing region of \SI{0.5}{\angstrom}. The potential is represented by a fully connected deep neural network with three layers of 240 neurons each. These are standard settings, which have also been used to study the phase diagram of water \cite{zhai_short_2023,bore_realistic_2023,sciortino2025constraints}. To train the MLIP, we built a training set by adversarial active learning implemented in the HyAL code~\cite{hyal} and documented in~\cite{cezar2025learning}. In brief, our adversarial active-learning approach is iterative: successive MLIPs are used to extend the training set by adding adversarial (high-error) structures, yielding robust and accurate MLIPs~\cite{cezar2025learning}. To ensure that the MLIP accurately describes interfaces and confined water under shear, not only bulk properties, the training set includes structures extracted from sheared ice--silica MD simulations. The complete training set comprises \num{1164} structures, including ice Ih, liquid water, liquid--ice coexistence, amorphous silica, silica--water, and sheared ice--silica interfaces, and is openly available in the associated data repository~\cite{data_repository}.

\subsection{System setup and sliding protocol}
\begin{figure}[!htbp]
    \centering \includegraphics[width=1\linewidth]{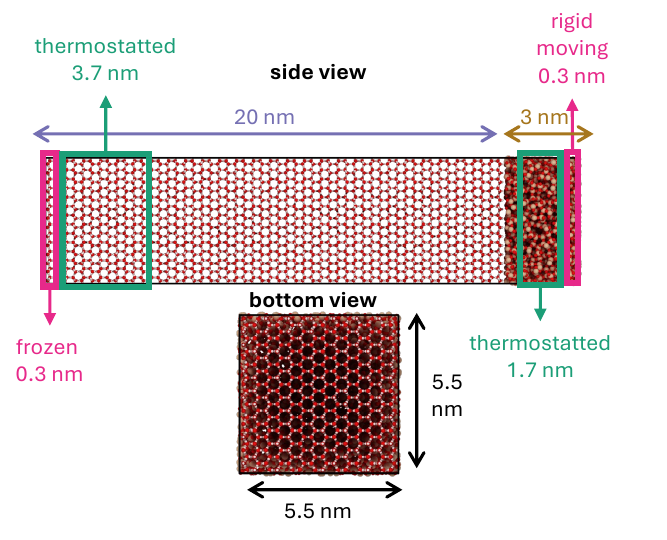}
    \caption{\textbf{Setup for nanoscale ice--glass simulations.}}
    \label{sm-fig:setup}
\end{figure}

    The initial simulation box shown in \figref{sm-fig:setup}  is prepared in three steps. First, a block of hexagonal ice is generated using GENICE to ensure the correct proton disorder~\cite{matsumoto2018genice}, with its cell geometry scaled to the equilibrium geometry at \SI{0.1}{\mega\pascal} and temperature $T$, with basal plane orientation at the surface. Second, we create an amorphous silica slab by melting beta-cristobalite at \SI{5000}{\kelvin}, deforming the box to perfectly match the ice equilibrium geometry's $L_{\text x}$ and $L_{\text y}$, and then quenching down from \SI{4000}{\kelvin} to $T$ with a quench rate of \SI{1E12}{\kelvin\per\second}, a rate that has been shown previously to yield converged amorphous silica structures \cite{erhardMachinelearnedInteratomicPotential2022}. Third, these two parts are combined with an offset of \SI{1}{\angstrom} to create an interface with a small gap in between. Finally, we perform equilibration under constant contact pressure $P_N$. To perform constant contact pressure simulations, we freeze \SI{3}{\angstrom} of the bottom ice layer and rigidify the top \SI{3}{\angstrom} of the top amorphous layer, constraining forces to the $z$-direction only, applying a force on each atom~\cite{heyes2011equivalence}:
\begin{equation}
    F_z=-\frac{P_{\text N}L_{\text x}L_{\text y}}{N_{\text{atoms,rigid}}}
\end{equation}
All data presented here were obtained from simulations carried out with \SI{10}{\mega\pascal} contact pressure. We maintain temperature by thermostatting \SI{37}{\angstrom} and \SI{17}{\angstrom} of the ice and silica, respectively, using a Langevin thermostat with a time constant of \SI{100}{\femto\second}~\cite{brunger1984stochastic}, correcting for the net center-of-mass motion contribution to the temperature. The thermostat is applied only to atoms far from the sliding interface, ensuring that it does not damp the sliding motion or bias the friction stress.
After equilibration of \SI{5}{\nano\second}, we apply a constant velocity to the rigid object in the x direction. By setting the forces to zero in the $x$ and $y$ directions, we maintain the sliding velocity in the $x$ direction.

\subsection{TIP4P/Ice + ClayFF runs}

To assess whether our findings depend on the specific atomistic model used to generate the premelting film, we repeated the ice--glass friction sweep with widely used non-reactive force fields: ClayFF~\cite{cygan2004molecular} for the amorphous silica and TIP4P/Ice~\cite{abascal2005potential} for water (rigid 3-site form, M-site implicit). All simulations were run with LAMMPS~\cite{thompson2022lammps}. The complete input and job scripts, initial structures, and analysis routines are openly available in the associated data repository~\cite{data_repository}; what follows is a high-level summary.

Because ClayFF is non-reactive, the silica slab cannot be melt-quenched in place. We instead inherit a frame from the equilibrated MLIP ice--silica interface, retype the atoms to ClayFF and TIP4P/Ice, and minimize and equilibrate under NVT in slab geometry, preserving the silanol coverage that emerged during the reactive MLIP equilibration. Production sliding runs are carried out in a fully periodic box with an $\approx \SI{30}{\angstrom}$ vacuum gap, a frozen bottom and a rigid-body top to which a constant lateral velocity and a constant downward force (matching the target contact pressure) are applied, and Langevin thermostats coupled to thin slabs just above the frozen layer and just below the rigid top. SHAKE~\cite{ryckaert1977numerical} is applied to the water O--H bond and H--O--H angle, allowing a \SI{2}{\femto\second} timestep. The runs are \SIrange{40}{100}{\nano\second} long, and the first \SI{15}{\nano\second} are discarded. The premelting-film thickness is measured as for the MLIP runs. The friction stress is the per-atom virial shear stress~\cite{thompson2009general} averaged over the ice and film between the two thermostatted slabs; in steady sliding the shear stress is uniform across this region, so no locked-stress correction is needed. Temperatures for this model are reported relative to its own melting point, \SI{270}{\kelvin}, rounded from the direct-coexistence value \SI{269.8(1)}{\kelvin} reported for TIP4P/Ice~\cite{conde2017melting}, the same method used for the DC-R$^2$SCAN melting point of \SI{289}{\kelvin}.

The production matrix consists of 45 conditions on the basal-plane (BP) ice orientation at \SI{10}{\mega\pascal}, with the full matrix and per-condition results tabulated in the data repository~\cite{data_repository}.

\section{MLIP validation}

\subsection{Training and held-out test-set parity}
\begin{figure*}[!htbp]
    \centering
    \includegraphics{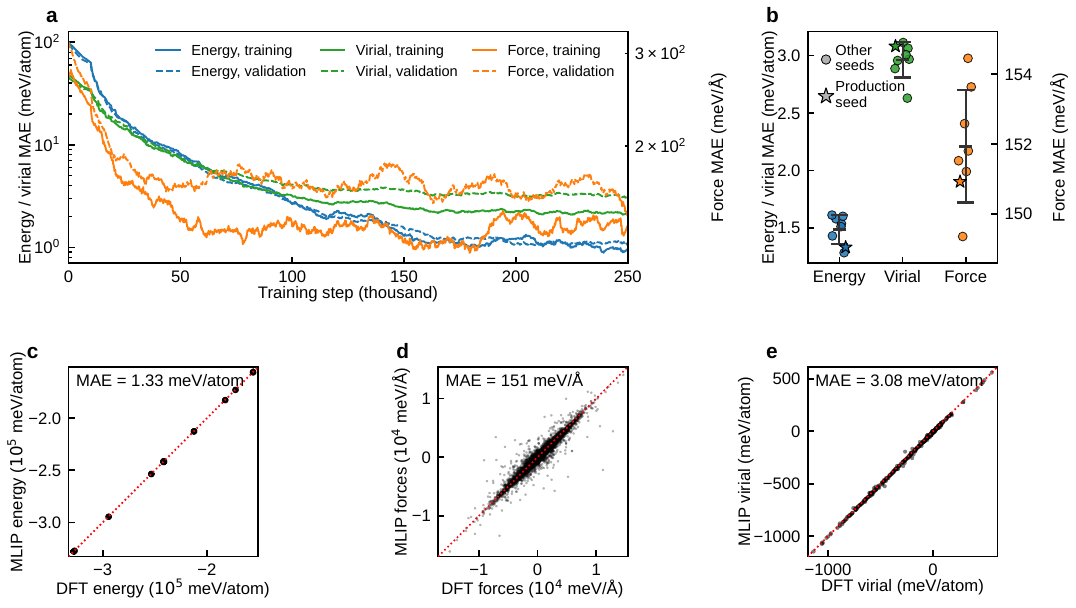}
    \caption{\textbf{Training and test-set performance of the DC-R\textsuperscript{2}SCAN MLIP.}
\textbf{a} Training (solid) and validation (dashed) MAE of energy, virial and
forces during the 250\,k optimisation steps of the production DeePMD-kit run.
Curves are rolling-averaged over 200 frames. Energy and virial share the
left axis (meV/atom); forces are read on the right axis (meV/\AA).
\textbf{b} Per-seed test-set MAE across eight MLIPs trained independently,
each with a different random seed governing the train/validation split and
the weight initialisation. Circles are the seven non-production seeds; the
star marks the production seed, with bars showing seed-to-seed mean $\pm$
standard deviation.
\textbf{c--e} Parity of DFT (CP2K) reference vs.\ MLIP prediction on the
held-out test set (116 configurations spanning ice, amorphous silica,
crystalline silica phases, liquid water and the silica--water interface):
\textbf{c} potential energy per atom, \textbf{d} force components, and
\textbf{e} virial per atom. Dotted red lines are $y = x$.}
    \label{sm-fig:mlip-accuracy}
\end{figure*}

Training and held-out test-set performance of the production MLIP is summarised in \figref{sm-fig:mlip-accuracy}: training and validation MAE curves stabilise within the \num{250}k-step optimisation budget (\figref{sm-fig:mlip-accuracy}a), per-seed test-set MAE is consistent across the eight independently trained MLIPs (\figref{sm-fig:mlip-accuracy}b), and parity against the DFT reference on the held-out test set is tight for energies, forces and virials (\figref{sm-fig:mlip-accuracy}c--e).

\subsection{Physical-property benchmarks}
\begin{figure*}[!htbp]
    \centering
    \includegraphics[width=1\linewidth]{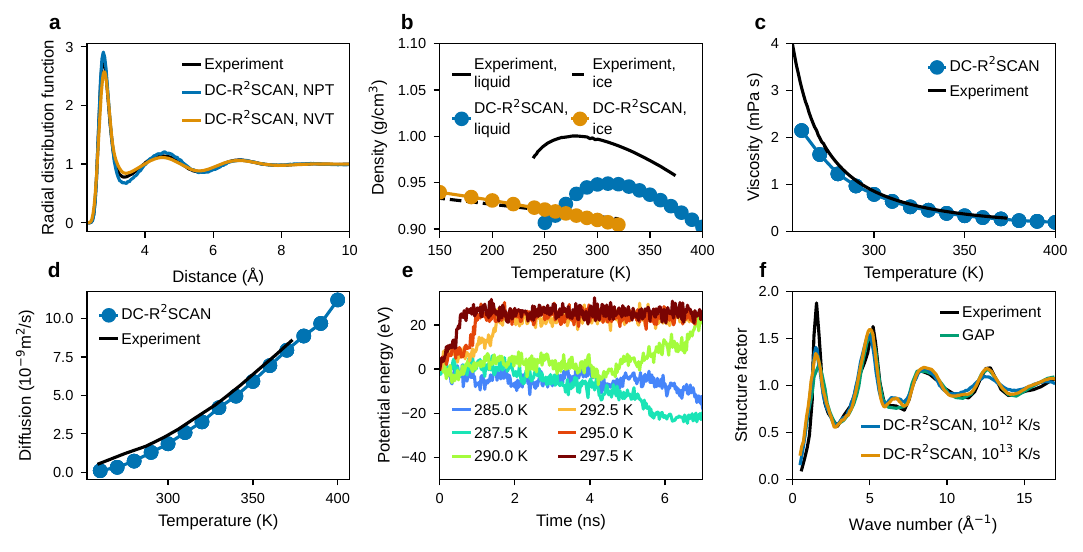}
    \caption{\textbf{Validation of DC-R$^2$SCAN MLIP for physical properties.} \textbf{a} Radial distribution function at \SI{300}{\kelvin} and \SI{0.1}{\mega\pascal} for experiment, and simulations using $NVT$ ensemble at experimental density, $NPT$ ensemble at \SI{0.1}{\mega\pascal}, \SI{300}{\kelvin}. \textbf{b} Density of liquid water and ice from experiments and simulations. \textbf{c} Potential energy for liquid--ice coexistence simulations at different temperatures. \textbf{d} Diffusion coefficient of liquid water at \SI{0.1}{\mega\pascal} for model and experiments~\cite{holz2000temperature,easteal1989diaphragm,mills1973self}. \textbf{e} Viscosity of liquid water at \SI{0.1}{\mega\pascal} for model and experiments~\cite{Lemmon2021,dehaoui2015viscosity}. \textbf{f} Structure factor of amorphous silica from experiments and GAP model~\cite{erhardMachinelearnedInteratomicPotential2022}, and simulations at different annealing rates.}
    \label{sm-fig:model-validation}
\end{figure*}

To further assess the accuracy of our DC-R$^2$SCAN MLIP, we validated it for physical properties of water and silica in \figref{sm-fig:model-validation}. The radial distribution function predicted by DC-R$^2$SCAN (\figref{sm-fig:model-validation}a) closely follows experiment, indicating an accurate liquid structure. The density isobars in \figref{sm-fig:model-validation}b show that the ice density is well reproduced, whereas the liquid-water density is underestimated by about \SI{5}{\percent}. In a separate publication~\cite{cezar2025learning}, we computed the water density for DC-R$^2$SCAN at \SI{0.1}{\mega\pascal} and \SI{300}{\kelvin} using an Allegro-based MLIP (\tabref{sm-tab:physical-properties}); the present value agrees within \SI{0.2}{\percent}, indicating well-converged training data. Direct coexistence (\figref{sm-fig:model-validation}c) yields a melting point of \SI{289}{\kelvin}, i.e.\ \SI{16}{\kelvin} above experiment. The amorphous-silica structure factor (\figref{sm-fig:model-validation}f) agrees well with experiment and a DFT-based GAP model for both quench rates~\cite{erhardMachinelearnedInteratomicPotential2022}, with a slight underestimation of the first peak.

\begin{table}[!htbp]
    \centering
    \caption{\textbf{Benchmark of DC-R$^2$SCAN MLIP.} All properties are reported for \SI{300}{\kelvin}, \SI{0.1}{\mega\pascal}, except for melting point. Ref.~\citenum{cezar2025learning} uses DC-R$^2$SCAN, but with the Allegro MLIP architecture and different training set structures~\cite{musaelian_learning_2023}.  }
    \label{sm-tab:physical-properties}
    \begin{tabular}{lcccc}
\toprule
Property &Unit & This work & Ref.\ \citenum{cezar2025learning} & Experiment \\
\midrule
Density &(\si{\gram\per\cubic\centi\metre}) & \num{0.948\pm 0.001} & \num{0.95\pm0.009} & \num{0.996} \\
    Melting point &(\si{\kelvin}) & \num{289\pm2} & --- & \num{273.15} \\
    Diffusion &($10^{-9}$\si{\square\metre\per\second}) & \num{1.8\pm0.2} & \num{1.59\pm0.09} & \num{2.4021} \\
    Viscosity& (\si{\milli\pascal\cdot\second}) & \num{0.7868} & --- & \num{0.8536} \\
\bottomrule
\end{tabular}
\end{table}

\section{Analysis protocols}

\subsection{Premelting-film thickness}
The premelting film thickness was estimated using the Polyhedral template matching~\cite{larsen2016robust} of OVITO~\cite{stukowski2009visualization}. Specifically, for each oxygen, we computed the Polyhedral template-matching RMSD for cubic and hexagonal diamond, and assigned it as ice-like if $0.01<\text{RMSD}<0.25$. Then, using a histogram and fitting the function $f(z)$:
\begin{equation}
    f(z)=\frac A2 \left[1-\tanh\left(\frac{z-z_{\text{ice}}}{k}\right)\right],
\end{equation}
we determine the position of the ice interface $z_{\text{ice}}$. Similarly, we make a histogram for the silica atoms and fit $g(z)$:
\begin{equation}
    g(z)=\frac A2 \left[1+\tanh\left(\frac{z-z_{\text{silica}}}{k}\right)\right],
\end{equation}
determining the position of the water--silica interface at $z_{\text{silica}}$. The premelting film thickness is then determined by $h=z_{\text{silica}}-z_{\text{ice}}$.

\subsection{Friction stress and effective viscosity}
The friction stress is given by:
\begin{equation}
    \sigma_{\rm f}=-\left<P_{xz}\right>_t,
\end{equation}
where $\left<P_{xz}\right>_t$ is the time average of the $xz$ component of the pressure tensor from the MD simulation. Because we freeze and rigidify the bottom and top parts of the system, we must account for the locked-stress contributions from these parts. We use the initial equilibration of \SI{5}{\nano\second}, during which there is no sliding motion and the friction stress should be zero, to estimate the locked stress. We then correct the friction stress time series and averages used in the main text by subtracting the locked stress from the measured stress. This correction works best for larger stresses that typically occur at higher sliding velocities, and is why we focus on sliding velocities above \SI{0.01}{\metre\per\second} in the main text. Finally, the viscosity of the premelting film is estimated using the relation:
\begin{equation}
    \eta=\frac{\sigma_{\rm f} h}{v}.
\end{equation}

\subsection{Steady-state averaging and raw time series}
\begin{figure*}[!htbp]
    \centering
    \includegraphics[width=1\linewidth]{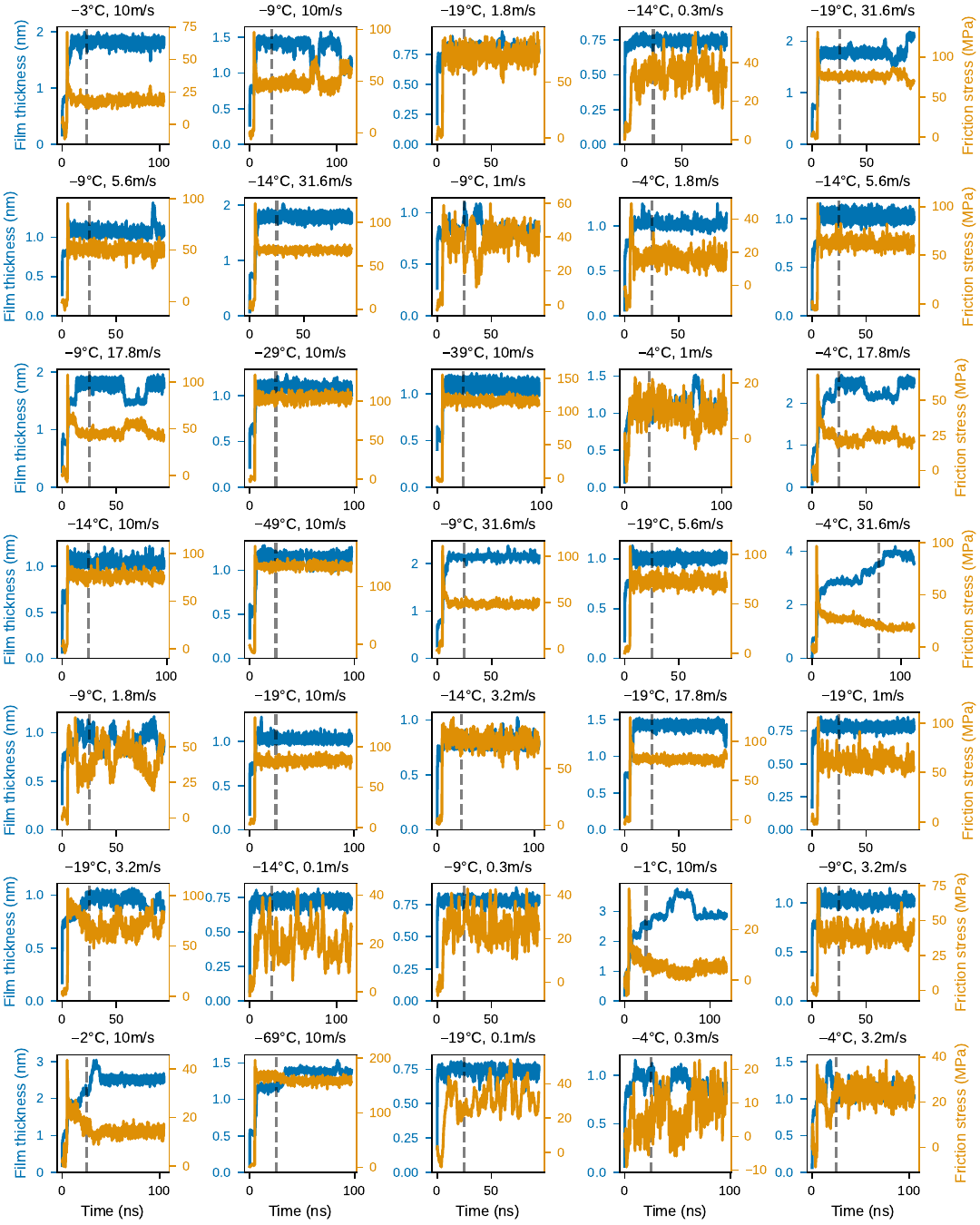}
    \caption{\textbf{Nanoscale ice friction time series 1.} Each time series is titled with contact temperature in \si{\celsius} with respect to melting point at \SI{289}{\kelvin} and sliding velocity in \si{\metre\per\second}. Dashed lines indicate the starting frame from which the averages of the steady state are estimated.
    }
    \label{sm-fig:series-1}
\end{figure*}

\begin{figure*}[!htbp]
    \centering
    \includegraphics[width=1\linewidth]{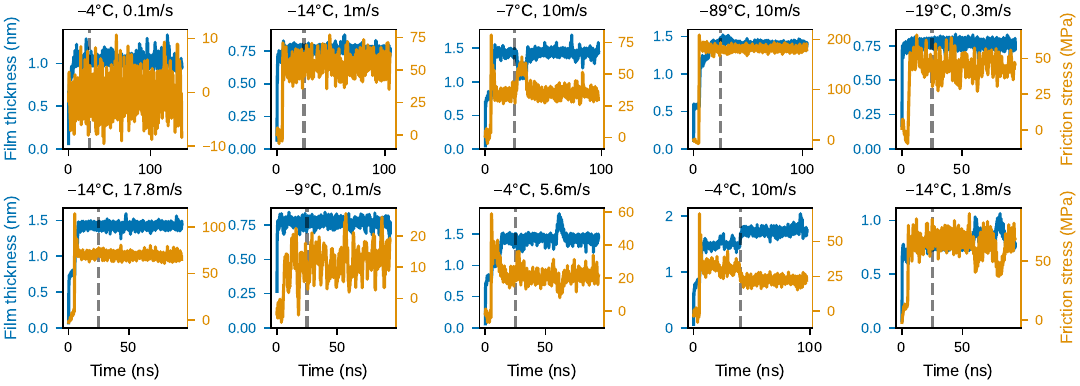}
    \caption{\textbf{Nanoscale ice friction time series 2.} Each time series is titled with contact temperature in \si{\celsius} with respect to melting point at \SI{289}{\kelvin} and sliding velocity in \si{\metre\per\second}. Dashed lines indicate the starting frame from which the averages of the steady state are estimated.
    }
    \label{sm-fig:series-2}
\end{figure*}

The raw time series for friction stress and premelting film thickness are shown in \figref{sm-fig:series-1} and \figref{sm-fig:series-2}. Averages for steady state are computed from the time series after an initial transient period. As shown in \figref{sm-fig:series-1} and \figref{sm-fig:series-2}, the premelting film grows from the initial interface upon the onset of sliding, reaching a steady state within nanoseconds. This initial time evolution is similar to that reported for displacement-driven amorphization by At{\i}la et al.~\cite{atila2025cold}. The steady-state film thickness and friction stress depend on contact temperature and sliding velocity, and serve as inputs to the macroscale frictional heating model.


\section{Supplementary analysis}

\subsection{Premelting-film structure}
\begin{figure}[!htbp]
    \centering
    \includegraphics[width=3.3in]{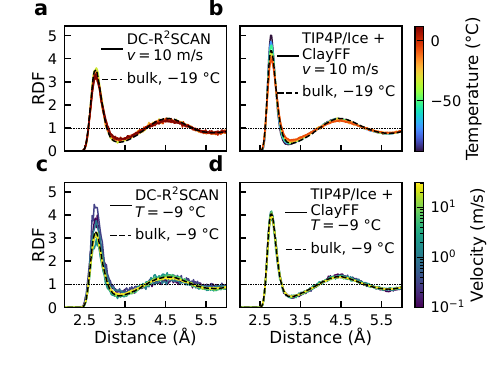}
    \caption{\textbf{Oxygen--oxygen RDF in the premelting-film midplane for both models.} \textbf{a, b} Temperature sweep at $v = \SI{10}{\metre\per\second}$: \textbf{(a)} DC-R$^2$SCAN, \textbf{(b)} ClayFF/TIP4P-Ice. \textbf{c, d} Velocity sweep at $T - T_\mathrm{m} = \SI{-9}{\celsius}$: \textbf{(c)} DC-R$^2$SCAN at \SI{280}{\kelvin}, \textbf{(d)} ClayFF/TIP4P-Ice at \SI{261}{\kelvin}. Strip half-width $= \max(\SI{0.5}{\angstrom}, 0.05 \times h)$; 500 frames averaged per condition.}
    \label{sm-fig:strip-rdf}
\end{figure}

We computed the oxygen--oxygen radial distribution function inside a thin strip centred on the midplane of the premelting film, counting only pairs in which both oxygens lie inside the strip and applying the slab-cap normalization so that $g(r) \to 1$ at large $r$. The film $g(r)$ matches the supercooled-water character that At{\i}la et al.\ identify for shear-amorphized ice~\cite{atila2025cold}, in both models and across all probed $(T, v)$ (\figref{sm-fig:strip-rdf}).

\subsection{Robustness of premelting-film thickness}

\begin{figure}[!htb]
    \centering
    \includegraphics[width=3.3in]{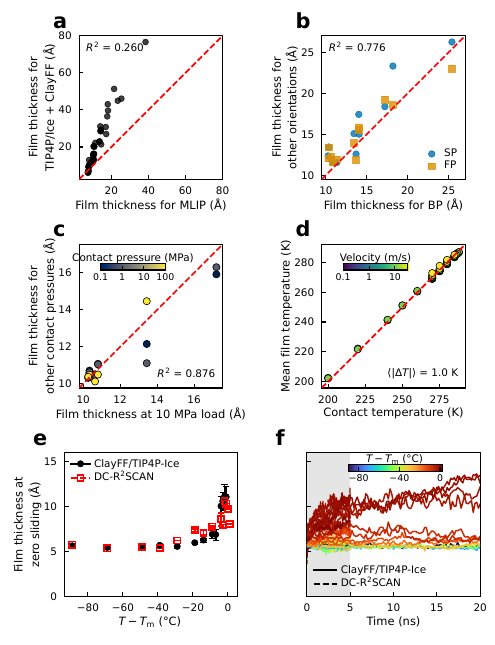}
    \caption{\textbf{Robustness of the premelting-film thickness} -- counterpart of Figure~3a--c for film thickness. \textbf{a} MLIP vs.\ ClayFF/TIP4P-Ice at matched $(v, \Delta T)$. \textbf{b} Basal vs.\ first/second prismatic (FP/SP) orientations at \SI{10}{\mega\pascal}. \textbf{c} Film thickness at \SI{10}{\mega\pascal} against other contact pressures (\SIrange{0.1}{100}{\mega\pascal}). \textbf{d} Mean temperature inside the film against the contact temperature; coloured by sliding velocity. \textbf{e, f} Premelting film with no sliding at all, in both models: \textbf{(e)} time-averaged thickness against $T - T_\mathrm{m}$, error bars are standard deviations; \textbf{(f)} the same runs as time series, coloured by $T - T_\mathrm{m}$. The shaded band in \textbf{(f)} marks the \SI{5}{\nano\second} DC-R$^2$SCAN window, so every dashed trace lies inside it and everything beyond it is ClayFF/TIP4P-Ice.}
    \label{sm-fig:thickness-parity}
\end{figure}

The same robustness analysis carried out for friction stress in the main text (\figref{fig:robustness}a--c) is shown for the premelting film thickness in \figref{sm-fig:thickness-parity}a--c. We also verify that the film sits at the temperature we index it by: the mean temperature inside the film deviates from the contact temperature by \SI{1.0}{\kelvin} on average (\figref{sm-fig:thickness-parity}d). In \figref{sm-fig:thickness-parity}e--f, we report the film thickness in the absence of any sliding motion: for ClayFF/TIP4P-Ice these are dedicated zero-velocity trajectories at thirteen temperatures, run for \SI{20}{\nano\second} at \SI{10}{\mega\pascal}; for DC-R$^2$SCAN they are the \SI{5}{\nano\second} zero-velocity segments preceding sliding in the production runs. In both models a premelting film is present for all the temperatures probed without any sliding.

\subsection{Extended sensitivity of upscaled friction coefficient}

\begin{figure*}[!t]
    \centering
    \includegraphics[width=1\linewidth]{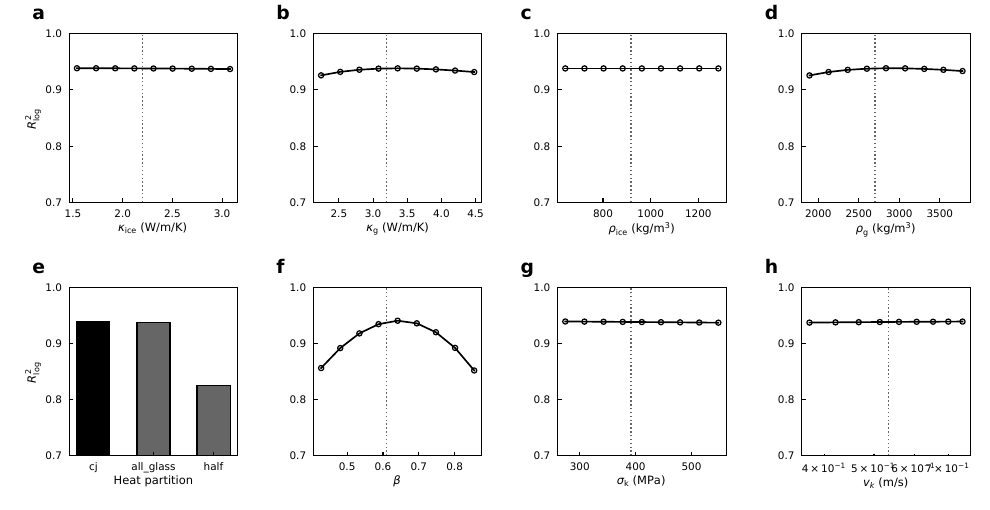}
    \caption{\textbf{Extended one-at-a-time sensitivity of the upscaled friction coefficient.} Each panel shows $R^2_\mathrm{log}$ of the upscaled friction coefficient against the pooled experimental dataset as one parameter is varied; the dotted line marks the default value. \textbf{a--d} Heat-model parameters $\kappa_\mathrm{ice}$, $\kappa_\mathrm{g}$, $\rho_\mathrm{ice}$, $\rho_\mathrm{g}$. \textbf{e} Heat-partition mode (Carslaw--Jaeger, all-glass, 50/50). \textbf{f--h} Nano fit parameters $\beta$, $\sigma_\mathrm{k}$, $v_\mathrm{k}$.}
    \label{sm-fig:extended-sensitivity}
\end{figure*}
\figref{fig:robustness}d--f in the main text reports the one-at-a-time (OAT) sensitivity of the upscaled friction coefficient to three model parameters that drive the largest variation in $R^2_{\log}$: the asperity radius $R_{\rm p}$, the low-velocity floor $\sigma_{\rm c}$, and the penetration hardness $\sigma_{\rm P}$. \figref{sm-fig:extended-sensitivity} reports the same OAT analysis for the remaining parameters of the macroscale heating pipeline---heat-model parameters $\kappa_{\rm ice}$, $\kappa_{\rm g}$, $\rho_{\rm ice}$, $\rho_{\rm g}$, the heat-partition mode, and nano fit parameters $\beta$, $\sigma_{\rm k}$, $v_{\rm k}$---confirming that $R^2_{\log}$ varies by less than 0.1 across $\pm 30\%$ around the default values of each.

\end{document}